\documentclass[aps,prl,twocolumn]{revtex4}

\usepackage{graphicx}% Include figure files

\usepackage{bm}% bold math
\usepackage{amsmath}
\usepackage{amssymb}

\newcommand{\fr}{\frac}
\newcommand{\sq}{\sqrt}
\newcommand{\lbl}{\label}

\newcommand{\beq}{\begin{equation}}
\newcommand{\eeq}{\end{equation}}
\newcommand{\beqar}{\begin{eqnarray*}}
\newcommand{\eeqar}{\end{eqnarray*}}

\newcommand{\sgn}{{\rm sgn}\,}

\newcommand{\ua}{\uparrow}
\newcommand{\da}{\downarrow}

\newcommand{\dx}{{\text d}}
\newcommand{\ex}{{\text e}}

\newcommand{\ix}{{\text i}}

\newcommand{\tx}{{\text t}}

\newcommand{\ch}{\hat{c}}

\newcommand{\hh}{\hat{h}}

\newcommand{\Hh}{\hat{H}}

\newcommand{\Jh}{\hat{J}}

\newcommand{\psih}{\hat{\psi}}

\newcommand{\Jc}{\mathcal{J}}

\newcommand{\Nc}{\mathcal{N}}

\newcommand{\sigb}{{\mbox{\boldmath{$\sigma$}}}}

\newcommand{\pd}{\partial}
\newcommand{\rtarr}{\rightarrow}

\newcommand{\Kb}{{\bf K}}

\newcommand{\Sb}{{\bf S}}

\newcommand{\ab}{{\bf a}}

\newcommand{\rb}{{\bf r}}

\newcommand{\At}{{\tilde{A}}}
\newcommand{\Bt}{{\tilde{B}}}

\newcommand{\dg}{\dagger}

\newcommand{\ran}{\rangle}

\newcommand{\al}{\alpha}
\newcommand{\be}{\beta}
\newcommand{\ga}{\gamma}

\newcommand{\de}{\delta}

\newcommand{\la}{\lambda}

\newcommand{\sig}{\sigma}

\newcommand{\vphi}{\varphi}
\newcommand{\eps}{\varepsilon}
\newcommand{\e}{\epsilon}
\newcommand{\lt}{\left}
\newcommand{\rt}{\right}
\newcommand{\Psih}{\hat{\Psi}}

\begin{document}
\title{Kondo effect in monolayer and bilayer graphene: physical realizations of the multi-channel Kondo models}
\author{Maxim Kharitonov and Gabriel Kotliar}
\affiliation{
Center for Materials Theory,
Rutgers University, Piscataway, New Jersey 08854, USA}
\date{\today}
\begin{abstract}

We perform a general group-theoretical study 
of the Kondo problem in monolayer and bilayer graphene around the charge neutrality point.
Utilizing the group representation theory, we derive from symmetry considerations a family of the Kondo models
for all symmetric placements with either 3- or 6-fold rotational axis of an impurity atom in an arbitrary orbital state.
We find six possible classes of the partially anisotropic four-channel Kondo model.
As the key result, we argue several possibilities to realize
the regime of the dominant channel-symmetric two-channel Kondo effect,
protected by the local symmetry and specifics of the graphene band structure.
Our findings open prospects for the observation of the rich multi-channel Kondo physics in graphene 
and the associated non-Fermi-liquid behavior.

\end{abstract}
\maketitle

\begin{figure}
$\mbox{\includegraphics[width=.25\textwidth]{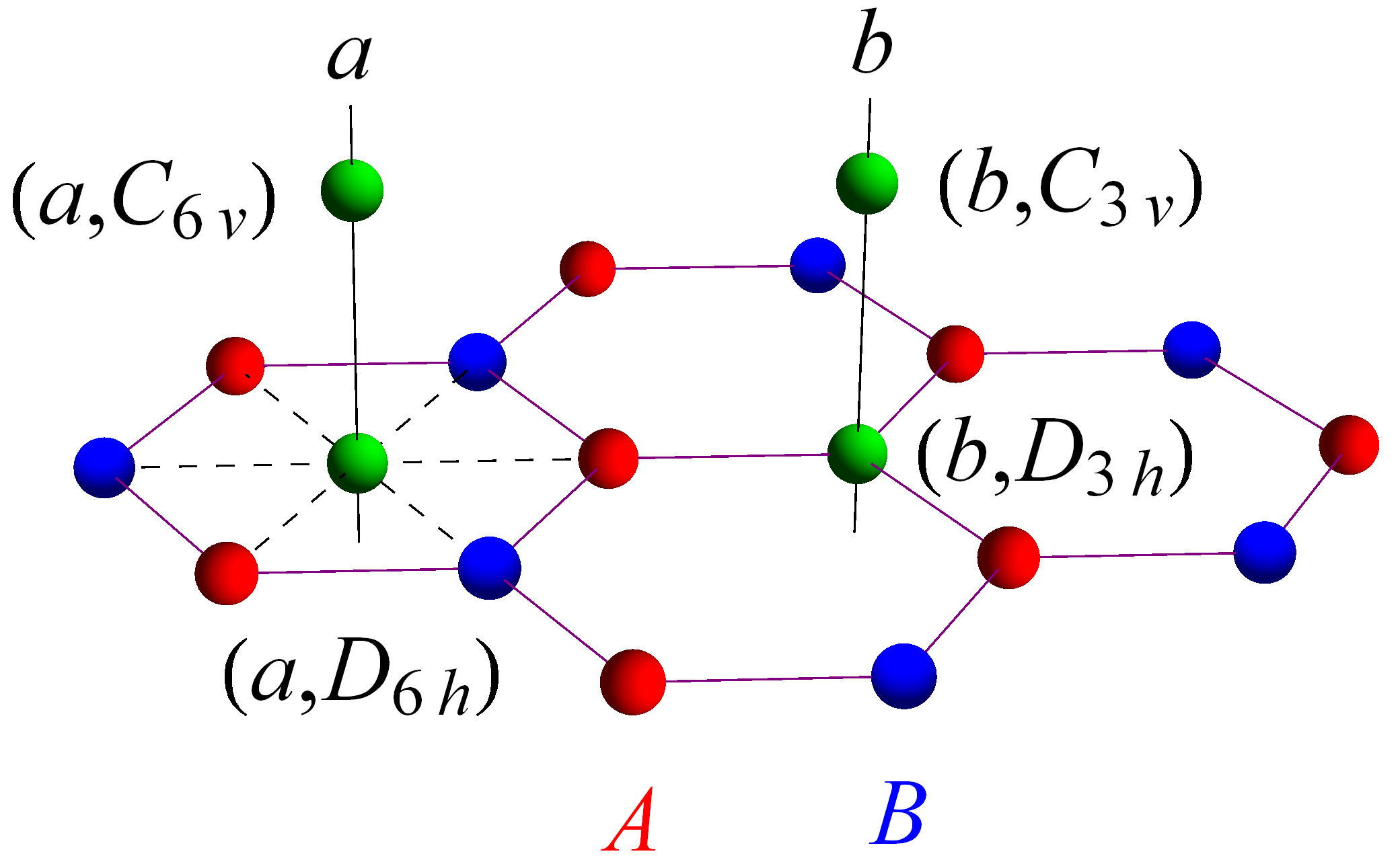}}\mbox{ }
\mbox{\includegraphics[width=.22\textwidth]{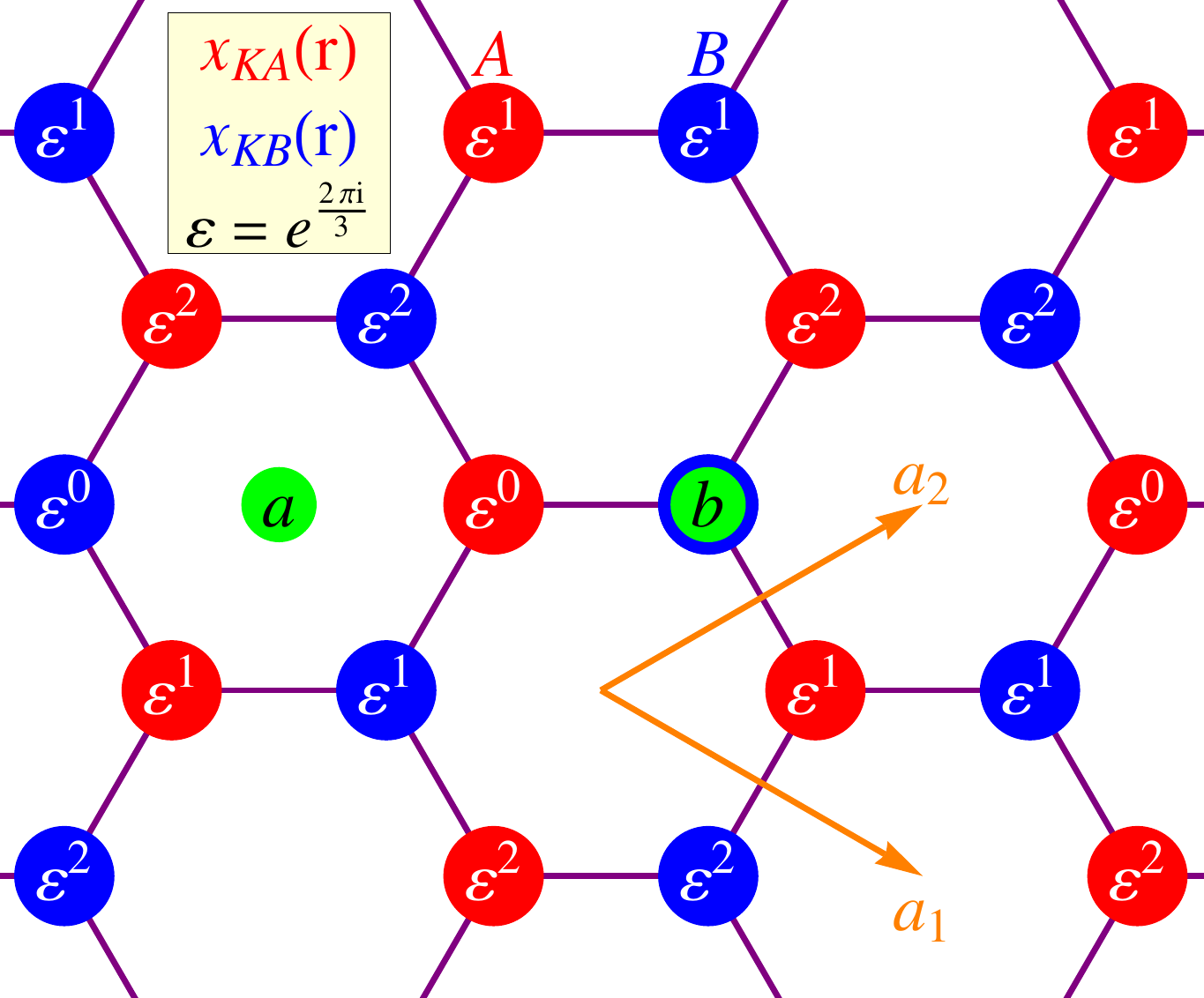}}$
$\mbox{\includegraphics[width=.25\textwidth]{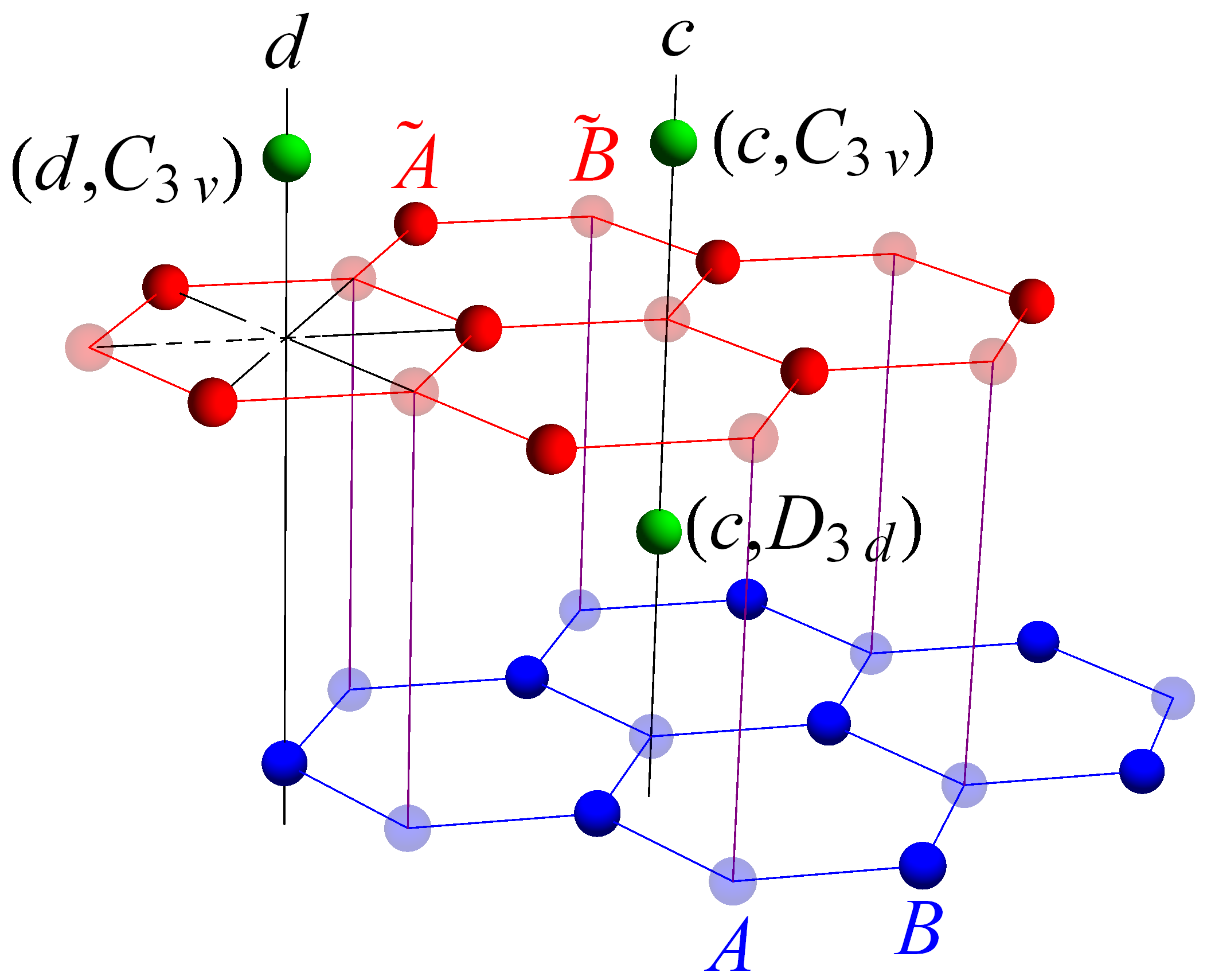}}\mbox{ }
\mbox{\includegraphics[width=.22\textwidth]{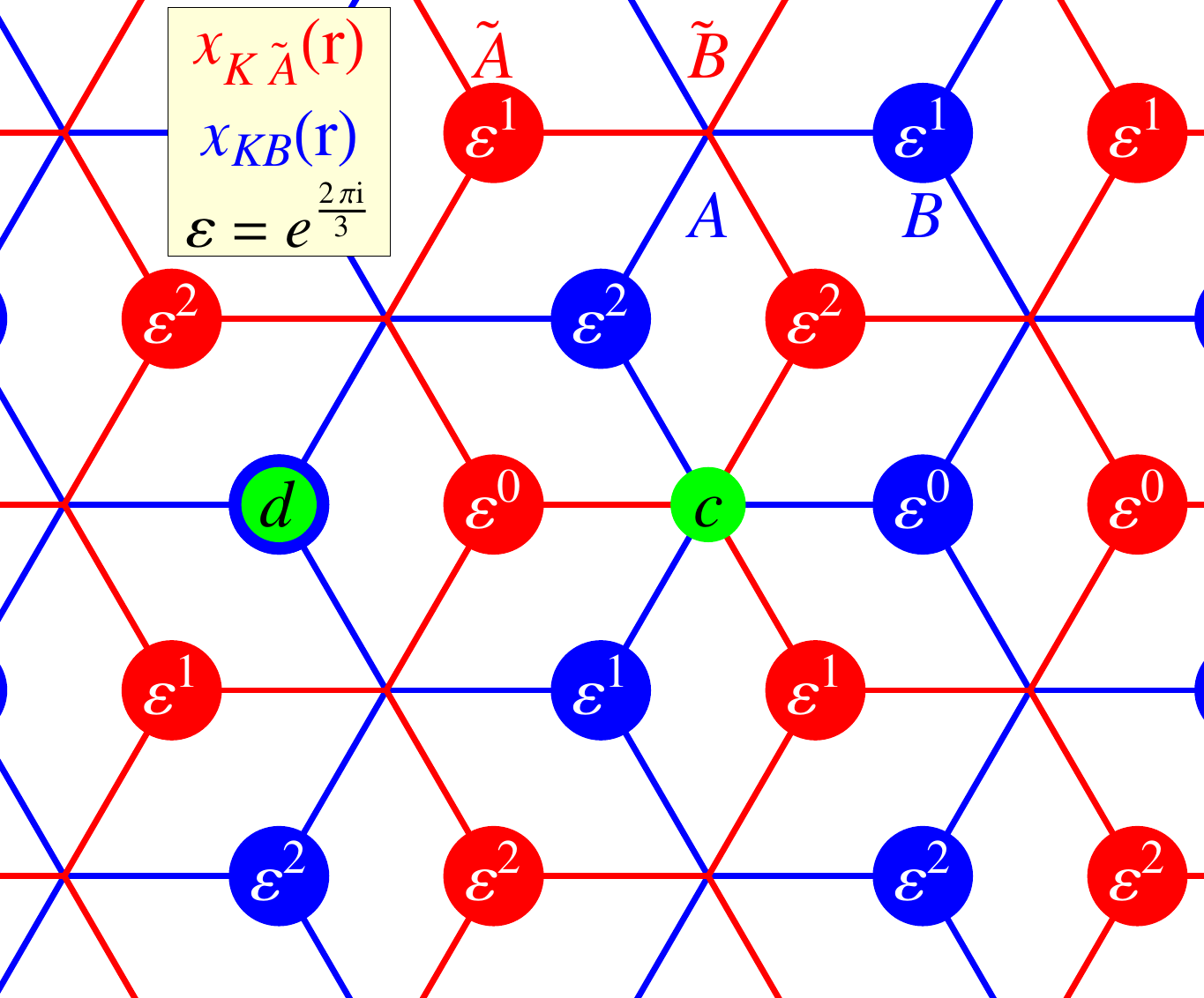}}$
\caption{(Color online) Lattice structure of MLG (top) and BLG (bottom) and considered impurity atom (IA) (green) placements
with either 3- or 6-fold rotational symmetry.
The phases $\eps^{0,1,2}$, $\eps=\ex^{2\pi\ix/3}$, of the Bloch wave-functions $\chi_{KA,KB}(\rb)$ in MLG (top right) and
$\chi_{K\At,KB}(\rb)$ in BLG (bottom right) are shown, $\chi_{K'\ldots}(\rb)=\chi_{K\ldots}^*(\rb)$.
}
\lbl{fig}
\end{figure}

\begin{table*}
\begin{tabular}{|l|c|c|c|}
    \hline
     &  $R_\ex$ & basis states of $R_\ex$ & squares of 2D IRs \\
    \hline
    MLG $(a,C_{6v})$ &
    $E_1+E_2$ &
$\chi_{E_1-} = \fr{1}{\sq2} (\chi_{KA} +\chi_{K'B})$,
$\chi_{E_1+} = \fr{1}{\sq2}(\chi_{KB} +\chi_{K'A})$, & $E_{1,2}\times E_{1,2}= A_1+A_2+E_1$\\
     $(a,D_{6h}=C_{6v}\times C_s)$ &   $E_1'' +E_2''$ &
$\chi_{E_2+} = \fr{1}{\sq2} (\chi_{K'A}-\chi_{KB})$,
$\chi_{E_2-} = \fr{1}{\sq2} (\chi_{KA}-\chi_{K'B})$ & $E_{1,2}^{',''}\times E_{1,2}^{',''}=A_1'+A_2'+E_1'$
\\\hline
      $(b,C_{3v})$ &
      $A_1+A_2+E$ &
$
    \chi_{E+}=\chi_{KA}$, $\chi_{E-}=\chi_{K'A}$ & $E\times E= A_1+A_2+E$
\\
      $(b,D_{3h}=C_{3v}\times C_s)$  & $A_1'' +A_2'' +E''$ &
$\chi_{A_1}=\fr{1}{\sq{2}}(\chi_{KB}+\chi_{K'B})$,
$\chi_{A_2}=\fr{1}{\sq{2}}(\chi_{KB}-\chi_{K'B})$, & $E^{',''}\times E^{',''}= A_1'+A_2'+E'$
\\
     \hline
     \hline
     BLG $(c,C_{3v})$ &  $2E=E^{\al}+E^{\be}$ &
$
    \chi_{E^\al+}\!=\!\chi_{K\At},\mbox{ }\chi_{E^\al-}\!=\!\chi_{K'\At},
$
$
    \chi_{E^\be+}\!=\!\chi_{K'B},\mbox{ }\chi_{E^\be-}\!=\!\chi_{KB}
$ & $E\times E= A_1+A_2+E$
\\\hline
      $(c,D_{3d}=C_{3v}\times C_i)$ &   $E_g+E_u$ &
    $\lt.\begin{array}{c}
        \chi_{E_u+} = \fr{1}{\sq2} (\chi_{K\At} +\chi_{K'B}),
        \chi_{E_u-} = \fr{1}{\sq2}(\chi_{K'\At} +\chi_{KB})\\
        \chi_{E_g+} = \fr{1}{\sq2} (\chi_{K\At} -\chi_{K'B}),
        \chi_{E_g-} = \fr{1}{\sq2}(\chi_{K'\At} -\chi_{KB})
    \end{array}\rt.$
& $E_{g,u}\times E_{g,u}= A_{1g}+A_{2g}+E_g$
\\\hline
      $(d,C_{3v})$ &   $A_1+A_2+E$ &
$    \chi_{E+}=\chi_{K'\At},\mbox{ }\chi_{E-}=\chi_{K\At},$ & $E\times E= A_1+A_2+E$
\\
     &   &
$
    \chi_{A_1}=\fr{1}{\sq2}(\chi_{KB}+\chi_{K'B}),
    \mbox{ }
    \chi_{A_2}=\fr{1}{\sq2}(\chi_{KB}-\chi_{K'B})
$ &
\\\hline
\end{tabular}
\caption{
Considered cases of impurity atom (IA) placement (first row), labeled by the rotational axis  $a,b,c,d$
and the local symmetry group $G$ (Fig.~\ref{fig}). Decomposition of the 4D space $R_\ex$ of the conduction electrons in terms of the IRs of $G$ (second row)
and the basis functions of IRs (third row).
Decomposition of the squares of 2D IRs (fourth row),
used for the construction of the exchange Hamiltonian $\Hh_J$ [Eq.~(\ref{eq:HJ}) and Tab.~\ref{tab:2}].
}
\lbl{tab:1}
\end{table*}

{\em Introduction and main results.}
The Kondo effect -- the interaction of the local spin or orbital degrees of freedom with the conduction electrons --
plays a fundamental role in a wide variety of condensed matter systems,
ranging from quantum dots to strongly correlated materials~\cite{NB,CZ}.
The Kondo effect in graphene~\cite{CNreview} has attracted significant interest~\cite{Sengupta,Hentschel,Dora,Uchoa1,Uchoa2,Uchoa3,Uchoa4,ZhuEPL,ZhuPRB,JacobKotliar,Zhuang,
Wehling1,Wehling2,Vojta,DellAnna,Cazalilla,FV,Lipinski,Chen,Brar,Andrei,Eelbo}
due to its peculiar electronic properties
and potential for the realization of the multichannel Kondo effects.
A number of theoretical studies of the quantum impurity problem
has been undertaken for monolayer graphene (MLG)~\cite{Sengupta,Hentschel,Dora,Uchoa1,Uchoa2,Uchoa3,Uchoa4,ZhuEPL,ZhuPRB,JacobKotliar,Zhuang,Wehling1,Wehling2,Vojta,DellAnna,Cazalilla,FV},
and much fewer for bilayer graphene (BLG)~\cite{Lipinski}.
However, the analysis of possible Kondo models in graphene has not yet been performed in its full generality.
In particular, the feasibility of the multichannel Kondo effect in graphene remained
a debated question:
both pro~\cite{Sengupta,ZhuEPL,DellAnna} and con~\cite{FV} arguments have been put forward.

Motivated by these interesting questions, 
in this Letter, we perform a general group-theoretical study of the Kondo effect in MLG and Bernal-stacked BLG
around the charge neutrality point (CNP).
Following the original recipe of Nozieres-Blandin~\cite{NB},
we utilize the group representation theory~\cite{H,LL3}
to derive the family of the four-channel Kondo models for all symmetric placements with either 3- ro 6-fold rotational axis
of an impurity atom (IA) in an arbitrary orbital state
without appealing to any microscopic details.
We find six possible classes of the Kondo models:
there are three cases for the structure of the conduction electron channels
and the impurity can be in either a singlet or doublet orbital state.

Most importantly, we argue that in several cases of the impurity placement 
MLG and  BLG band structure allows for the realization of the
%[channel-symmetric]
two-channel Kondo effect,
where the exchange couplings
for the pair of equivalent channels $E_\ex \pm$ belonging to one two-dimensional (2D) irreducible representation (IR) $E_\ex$
are dominant. In these cases, in the low-energy regime,
the symmetric two-channel Kondo model
\beq
    \tilde{\Hh}_J^\text{sglt}=\psih_{E_\ex}^\dg(0) (J^{\rho E_\ex E_\ex}+J^{\sig E_\ex E_\ex}\sigb\Sb)\psih_{E_\ex}(0)
\lbl{eq:HJ1low}
\eeq
is realized in the orbital singlet case and the two-channel Kondo model
\beq
    \tilde{\Hh}_J^\text{dblt}=\psih_{E_\ex}^\dg(0) \sum_{\ga=0,x,y,z}(J^{\rho E_\ex E_\ex}_\ga+J^{\sig E_\ex E_\ex}_\ga\sigb\Sb)\tau_\ga T_\ga \psih_{E_\ex}(0)
\lbl{eq:HJ2low}
\eeq
with partially anisotropic ($J^{\ldots}_x=J^{\ldots}_y\neq J^{\ldots}_z$) orbital Kondo interactions is realized in the orbital doublet case
Here, $\psih_{E_\ex}=(\psih_{E_\ex+\ua},\psih_{E_\ex+\da},\psih_{E_\ex-\ua},\psih_{E_\ex-\da})^\tx$
is a spinor in the product of the channel ($E_\ex \pm$) and spin ($\ua,\da$) spaces,
$\tau_\ga$ and $T_\ga$ are the unity ($\ga=0$) and Pauli ($\ga=x,y,z$) matrices in the channel and impurity orbital doublet spaces,
respectively,
$\sigb=(\sig_x,\sig_y,\sig_z)$ are the spin Pauli matrices of the conduction electrons,
and $\Sb=(S_x,S_y,S_z)$ are the impurity spin operators.

Both models present considerable physical interest, largely due to the non-Fermi-liquid behavior exhibited in a number of regimes,
but have proven challenging to realize in practice.
Our findings thus pose MLG and BLG as promising materials
for the realization of the rich multichannel Kondo physics
and the associated non-Fermi-liquid behavior.

{\em Graphene lattice and local symmetry of the impurity atom.}
MLG and BLG~\cite{CNreview} are two-dimensional carbon allotropes, shown in Fig.~\ref{fig}.
Both have a triangular Bravais lattice with the primitive translation vectors $\ab_{1,2}$.
MLG has $P6/mmm$ ($D_{6h}^1$, $\#191$, hexagonal system) space group and $D_{6h}$ point group;
its unit cell contains two atoms denoted $A$ and $B$.
BLG has $P\bar{3}m1$ ($D_{3d}^3$, $\#164$, trigonal system) space group and  $D_{3d}$ point group;
its unit cell contains four atoms $A$, $B$, $\At$, and $\Bt$, two in each layer.

In the presence of the IA,
the spatial symmetry is reduced to a point group $G$,~\cite{H,LL3}
with the center of the impurity being the fixed point.
For the realization of the channel-symmetric 2-channel Kondo effect,
it is necessary that $G$ has 2D IRs,
which is the case for hexagonal lattice if $G$ contains either 3- or 6-fold rotational axis.
Lower-symmetry point groups have no 2D IRs and the symmetric two-channel Kondo model is not feasible in these cases.
Therefore, in this paper, we consider only such impurity placements that $G$ contains either 3- or 6-fold
rotational axis (these, at the same time, may be the likely adsorption sites).

For MLG, there are two such vertical axes, denoted $a$ and $b$ (Fig.~\ref{fig}).
For the axis $a$ going through the center of the carbon hexagon,
$G=C_{6v}$, if the IA is out of the MLG $z\leftrightarrow -z$ mirror plane
and $G=D_{6h}=C_{6v}\times C_s$ (equal to the point group of MLG), if the impurity center is in the mirror plane
(here, $C_s$ is the group of mirror reflection only).
For the axis $b$ going through the center of the carbon atom,
$G=C_{3v}$, if the IA is out of the MLG plane, and $G=D_{3h}=C_{3v}\times C_s$, if the IA is in the MLG plane
(i.e., IA substitutes the carbon atom).

For BLG, there also are two such vertical axes, denoted as $c$ and $d$ (Fig.~\ref{fig}).
For the axis $c$ going through the center of the hexagon of one layer and a carbon atom ($B$) of the other layer,
$G=C_{3v}$ for any of the vertical position of IA on the axis.
For the axis $d$ going through two carbon atoms ($A$ and $\Bt$) of the layers,
$G=D_{3d}=C_{3v}\times C_i$ (equal to the BLG point group, $C_i$ is the point group of inversion only),
if the IA is at the midpoint between the layers,
and $G=C_{3v}$ for any other IA placement on the axis.
These cases are summarized in Tab.~\ref{tab:1}.

{\em Impurity degrees of freedom.}
A generalized Kondo model~\cite{NB,CZ}
describes minimal coupling of the conduction electrons to the spin and, if present,  orbital degrees of freedom
of ground state of a decoupled IA, i.e., neglecting the hybridization to the conduction states.
The decoupled IA has a definite electron occupancy number $N$, spin $S$
(according to the Hund's rule, $S=N/2$ and $S=2l+1-N/2$ for less- and more-than-half-filled orbital with angular momentum $l$, respectively).
In the crystalline environment of the lattice,
the orbital ground state of IA belongs to one of the IRs $R_\ix$ of the local symmetry group $G$ (we assume spin-orbit interactions weak).
In all cases we consider (Tab.~\ref{tab:1}), $G$ has only 1D or 2D IRs~\cite{H,LL3}.
Thus, the IA ground state is either an orbital {\em singlet} $|S_z\ran$
or {\em doublet} $|\al\ran\otimes|S_z\ran$ if it belongs to one of the 1D ($R_\ix=A_\ix$) or 2D ($R_\ix=E_\ix$) IRs, respectively.
Here, $A_\ix$ and $E_\ix$ denote {\em arbitrary} 1D and 2D IRs of the impurity orbital state, respectively,
$S_z=-S,\ldots,S$, and $\al=\pm$ are the quantum numbers of the orbital doublet $R_\ix=E_\ix$.
We choose the basis states $|\al=\pm\ran$ so that they transform as $\ex^{\al\ix\vphi}$ under $C_{3v}$
if $R_\ix=E$, and as $\ex^{\al\ix\vphi}$ and $\ex^{-\al2\ix\vphi}$ under $C_{6v}$ if $R_\ix=E_2,E_1$, respectively.

Thus, among the variety of microscopic possibilities of different $l$, $N$, and sequences of crystal field splittings,
group-theoretically, there are only two different classes of the orbital state of the IA.

{\em Conduction electrons in MLG and BLG.}
The electronic band structure in both MLG and BLG~\cite{Wallace,MF,CNreview}
around zero doping is governed by four Bloch states $\chi_\mu(\rb)$ (Fig.~\ref{fig}),
$\mu=KA,KB,K'A,K'B$ in MLG and $\mu=K\At,KB,K'\At,K'B$ in BLG,
at two high-symmetry points $\Kb = \fr{4\pi}{3 |\ab_1|^2} (\ab_1-\ab_2)$ and $\Kb'=-\Kb$ in the Brillouin zone, referred to as valleys $K$ and $K'$,
with energy $\e=0$ exactly at the charge neutrality point (CNP).
The states in two valleys are related by the time reversal symmetry: $\chi_{K'\ldots}(\rb)=\chi_{K\ldots}^*(\rb)$.
In MLG, the two states per valley reside on either $A$ or $B$ sublattice and are labeled accordingly.
In BLG, the two states per valley reside on either $\At$ or $B$ sublattice,
located in different layers, while their weight on $A$ and $\Bt$ sublattices vanishes.
These four states $\chi_\mu(\rb)$ form  4D IRs of the respective space groups of MLG and BLG.

The vicinity of the CNP can then be described utilizing the $k\cdot p$-theory expansion of the electron field operator
\beq
    \Psih_\sig^\text{MLG,BLG}(\rb)= \sum_\mu\chi_\mu(\rb) \psih_{\mu\sig}(\rb)
\lbl{eq:Psi}
\eeq
in terms of the exact Bloch wave-functions at $\e=0$ and the operators $\psih_{\mu\sig}(\rb)$ that vary over large spatial scales
($\sig=\ua,\da$ is the spin projection).
In one valley, the electron motion in MLG and BLG is described
by the chiral linear (massless Dirac equation)
and quadratic Hamiltonians in the sublattice space, respectively,
\[
    \hh_K^\text{MLG}= v\lt(\begin{array}{cc}  0 & p_- \\ p_+ & 0 \end{array}\rt)_{AB},
    \mbox{ }
    \hh_K^\text{BLG}= \fr{1}{2m}\lt(\begin{array}{cc}  0 & p_+^2 \\ p_-^2 & 0 \end{array}\rt)_{\At B}.
\]
Here, $p_\pm = p_x\pm\ix p_y$, $p_{x,y}= -\ix\pd_{x,y}$, $v\approx10^8\text{cm}/\text{s}$, and $m\approx m_e/20$.
The Hamiltonian in the other valley is the time-reversal counterpart, $\hh_{K'}^{\ldots}=\hh_K^{\ldots*}$.
These two-band models are valid at energies $\e\ll t$ bellow the nearest neighbor hopping amplitude $t\approx 3\text{eV}$ in MLG
and $\e \ll t_\perp$ below the interlayer hopping amplitude $t_\perp \approx 0.3\text{eV}$ in BLG.

{\em Classifying the conduction electrons.}
To proceed with the derivation of the Kondo models, we classify the conduction electron states according to the local symmetry group $G$.
It is sufficient to consider the states exactly at the CNP, treating $\psih_{\mu\sig}(\rb)$ as constants,
in which case $\Psih^{\text{MLG},\text{BLG}}(\rb)$ form a 4D representation $R_\ex$ of $G$.
Applying the symmetry operations of $G$ to $\chi_\mu(\rb)$ in each considered case,
we obtain the decomposition of $R_\ex$ into the IRs of $G$ and their basis functions (second and third rows of Tab.~\ref{tab:1}).
(In doing so, we keep in mind that the $p_z$ atomic orbitals of $\chi_\mu(\rb)$
have odd parity under mirror reflection in the graphene plane and inversion.)
Similarly to the impurity states $|\al=\pm\ran$, we choose the basis states of the 2D IRs
so that they transform as $\chi_{E\pm}\sim\ex^{\pm\ix\vphi}$ under $C_{3v}$ and
as $\chi_{E_2\pm}\sim\ex^{\pm\ix\vphi}$ and $\chi_{E_1\pm}\sim\ex^{\mp2\ix\vphi}$ under $C_{6v}$.
This convention eventually leads to the most natural form of the orbital Kondo interactions (Tab.~\ref{tab:2}).
The electron field operators can be rewritten in the new bases as $\Psih(\rb)^\text{MLG,BLG}=\sum_\eta \chi_\eta(\rb)\psih_{\eta\sig}(\rb)$,
where $\eta$ label the basis states of the IRs, e.g., $\eta=A_1,A_2,E+,E-$ in the MLG $(b,C_{3v})$ case.
For further use, we arrange the operators $\psih_{\eta\sig}$ into the eight-component spinors $\psih$ presented in Tab.~\ref{tab:2},
where $\psih_\eta=(\psih_{\eta\ua},\psih_{\eta\da})^\tx$.
For the cases $(a,C_{6v})$ and $(b,C_{3v})$ in MLG,
our results for the classification of $\psih_\eta$ agree with the earlier results of Refs.~\cite{Wehling1,Wehling2,ZhuEPL}.

We find that the seven considered cases of the impurity placement fall into 3 different classes, presented as columns of Tab.~\ref{tab:2}:
the conduction states $R_\ex$ split into
(I) two different 2D IRs;
(II) one 2D and two 1D IRs;
(III) two 2D IRs of the same type.

{\em Four conduction channels.}
At the length scale of the slowly varying operators $\psih_{\mu\sig}(\rb)$,
the atomic impurity may be considered as a point object, whose degrees of freedom couple only to the conduction states
with nonvanishing weight $\psih_{\mu\sig}(0)$ at its position $\rb=0$.
As the solution of the eigenvalue problem for $\hh_{K,K'}^\text{MLG,BLG}$ in polar coordinates shows,
see Refs.~\cite{ZhuEPL,Sengupta,DellAnna} and Supplementary Material,
for both MLG and BLG there is exactly one such radial channel for each of the four components $\mu$ of $\psih_{\mu\sig}(\rb)$.
By the standard ``unfolding'' procedure~\cite{CZ}, where the outgoing and incoming radial waves are mapped to the plane waves in $s>0$ and $s<0$
regions of the effective 1D axis, respectively, these four channels can be represented as chiral 1D channels with the kinetic energy
\beq
    \Hh_0=\int \dx s\, \psih^\dg(s) (-\ix\pd_s-\e_F) \psih(s).
\label{eq:H0}
\eeq
We will preserve the same notation for the 1D fields $\psih(s)$ as for the 2D $\psih(\rb)$ ones.
Thus, for an atomic-size impurity, a 4-channel Kondo model is generically realized in MLG~\cite{Sengupta,DellAnna,ZhuEPL} and BLG around the CNP.

\begin{table*}
\begin{tabular}{|c|c|c|c|}
    \hline
     $\hh_J^{\rho,\sig}$ & (I) MLG: $(a,C_{6v}/D_{6h})$; BLG: $(c,D_{3d})$  & (II) MLG: $(b,C_{3v}/D_{3h})$; BLG: $(d,C_{3v})$ & (III) BLG: $(c,C_{3v})$ \\
    \hline
    sglt.: &
$
    \lt(\begin{array}{cc}J^{E_{\ex1}E_{\ex1}} & 0\\ 0  & J^{E_{\ex2} E_{\ex2}}  \end{array}\rt)
        \otimes \tau_0
$
     &
$
  \lt(\begin{array}{cccc}
    J^{A_1A_1} & 0 & (0,0) \\
    0& J^{A_2A_2} &  (0,0)\\
    (0,0)^\tx  & (0,0)^\tx &   J^{EE}\tau_0
         \end{array}\rt)
$
    &
$

       \lt(\begin{array}{cc}J^{E^\al E^\al} & J^{E^\al E^\be}  \\ J^{E^\al E^\be*}  & J^{E^\be E^\be}  \end{array}\rt)
        \otimes \tau_0
$
\\
    dblt.: &
$
  \sum_\ga \lt(\begin{array}{cc}J_\ga^{E_{\ex1} E_{\ex1}} & 0\\ 0  & J_\ga^{E_{\ex2} E_{\ex2}}  \end{array}\rt)
        \otimes \tau_\ga T_\ga
$
    &
$
    \lt(\begin{array}{cccc}
    J^{A_1A_1} T_0 & J^{A_1A_2} T_z &  J^{A_1E}(T_-,T_+)\\
    \text{h.c.}& J^{A_2A_2} T_0 &  J^{A_2E}(T_-,-T_+)\\
    \text{h.c.}  & \text{h.c.} &   \sum_\ga J^{EE}_\ga \tau_\ga T_\ga
         \end{array}\rt)
$
    &
$
    \sum_\ga \lt(\begin{array}{cc}J_\ga^{E^\al E^\al} & J_\ga^{E^\al E^\be}  \\ J_\ga^{E^\al E^\be*}  & J_\ga^{E^\be E^\be}  \end{array}\rt)
        \otimes \tau_\ga T_\ga
$
  \\\hline
 & $\psih=(\psih_{E_{\ex1}+},\psih_{E_{\ex1}-},\psih_{E_{\ex2}+},\psih_{E_{\ex2}-})^\tx$
 & $\psih=(\psih_{A_1},\psih_{A_2},\psih_{E+},\psih_{E-})^\tx$
 & $\psih=(\psih_{E^\al+},\psih_{E^\al-},\psih_{E^\be+},\psih_{E^\be-})^\tx$
    \\\hline
\end{tabular}
\caption{
The most general forms allowed by symmetry of the exchange interaction Hamiltonian  $\Hh_J$ [Eq.~(\ref{eq:HJ})]
in MLG and BLG around the CNP for all possible symmetric placements with either 3- or 6-fold rotational axis of the impurity atom
(Fig.~\ref{fig} and Tab.~\ref{tab:1}).
The entries are the expressions for $\Jh^{\rho,\sig}$; both have identical structure in $R_\ex$
but their own sets of coupling constants $J^{\rho \ldots}_{\ldots}$ and $J^{\sig \ldots}_{\ldots}$,
we suppress the indices $\rho,\sig$ for brevity.
There are three classes, (I), (II), and (III) (columns), of the conduction electron orbital states $R_\ex$
and two classes, singlet (sglt., $R_\ix=A_\ix$) and doublet (dblt., $R_\ix=E_\ix$) (rows), of the impurity orbital state $R_\ix$.
In (I), $E_{\ex1,\ex2}=E_{1,2},E_{1,2}'',E_{g,u}$; in (II), $T_\pm=T_x\pm\ix T_y$.
In the orbital doublet case, the summation goes over $\ga=0,x,y,z$
and everywhere $J^{\ldots}_x=J^{\ldots}_y$, since $\tau_x T_x+\tau_y T_y$ is an invariant.
}
\lbl{tab:2}
\end{table*}

{\em Kondo models from the symmetry approach.}
We now derive the Kondo models.
As first outlined by Nozieres and Blandin~\cite{NB} and later implemented in a number of works for various Kondo systems~\cite{Toth,CZ},
the most general possible form of the exchange interaction Hamiltonian $\Hh_J$
can be efficiently obtained based on the symmetry grounds without appealing to any microscopic model:
$\Hh_J$ must remain invariant under all symmetry operations of the system.
The most general form invariant under spin rotations reads
\beq
    \Hh_J=\psih^\dg(0) (\Jh^\rho+\Jh^\sig \sigb \Sb)\psih(0).
\lbl{eq:HJ}
\eeq

In Eq.~(\ref{eq:HJ}), $\Jh^{\rho,\sig}$ are the operators in the orbital sector $R_\ex\otimes R_\ix$:
they are $4\times4$ matrices in the orbital space $R_\ex$ of the conduction electrons
and, in the impurity doublet case,
they also contain the orbital ``isospin'' operators $T_\ga$, $\ga=0,x,y,z$ (unity and Pauli matrices)
acting in the space $R_\ix=E_\ix$ of the states $|\al=\pm\ran$.

The operators $\Jh^{\rho,\sig}$ must remain invariant under the orbital symmetry group $G$~\cite{TR}.
Such invariant form is efficiently constructed using the algebra of the group representation theory~\cite{H,LL3} as follows~\cite{NB,CZ}.
The operators
\beq
    \Jh^{\rho,\sig} \sim (R_\ex \times R_\ex^\dg )\times (R_\ix\times R_\ix^\dg)
\lbl{eq:P}
\eeq
transform as a product of four representations of $G$, where $R_\ex\times R_\ex^\dg$  and $R_\ix\times R_\ix^\dg$
describe the transformation properties in the conduction electron and impurity subspaces, respectively.
The decomposition of the product (\ref{eq:P}) into IRs can readily be calculated~\cite{H,LL3}.
The only allowed terms in $\Jh^{\rho,\sig}$
are the invariants, which transform according to the unity IR $A_1/A_1'/A_{1g}$ of $G$;
each invariant may enter $\Jh^{\rho,\sig}$ with its own coupling constant $J$.
The explicit form of these invariants is constructed by utilizing the transformation properties of $\psi_{\eta\sig}$ and $|\al=\pm\ran$ under $G$.

This procedure yields the most general forms allowed by symmetry
of the Kondo exchange interaction Hamiltonians $\Hh_J$ [Eq.~(\ref{eq:HJ})], presented in Tab.~\ref{tab:2};
details of the derivation are provided in the Supplementary Material.
The contents of Tab.~\ref{tab:2}, along  with Tab.~\ref{tab:1} and Eqs. (\ref{eq:H0}) and (\ref{eq:HJ}), constitute the central result of our work.
They describe the family of the four-channel Kondo models
\[
    \Hh=\Hh_0+\Hh_J
\]
in MLG and BLG in the vicinity of the CNP
for all possible seven cases (Tab.~\ref{tab:1})
of the symmetric placements of the IA with either 3 or 6-fold rotational axis.
In the rest of the paper, we discuss their key properties and physical implications.

{\em Main properties.}
We find 6 possible classes of the Kondo models:
there are 3 above-mentioned classes (I), (II), and (III) (columns of Tab.~\ref{tab:2})
for the conduction electron states
and two classes of the impurity orbital state, singlet (sglt., $R_\ix=A_\ix$) and doublet (dblt., $R_\ix=E_\ix$) (rows of Tab.~\ref{tab:2}).
All cases of the impurity placement within one class have identical structure of the Kondo model.

In the orbital-singlet class (I) and (II) models,
when the conduction states $R_\ex$ break into IRs of different type,
the channels are not coupled by the exchange interaction and the two channels belonging to the same 2D IR
are characterized by the same coupling constant, as protected by symmetry.
These are prerequisites for the realization of the two-channel Kondo effect.
On the other hand, in the orbital-singlet class (III) model,
the conduction sea $R_\ex=2E=E^\al+E^\be$ consists of two 2D IRs of the same type $E$
(the labels $\al,\be$ are used to distinguish between the two subspaces).
As a result the ``conversion'' processes
$E^\al\leftrightarrow E^\be$, whereby the conduction electrons are transferred between two IRs, are present.

In the orbital doublet case, for any class (I), (II), or (III), the two channels belonging to each 2D IR $E_\ex$
couple to the impurity via anisotropic orbital Kondo interaction
$J_0^{\ldots} \tau_0 T_0+J_z^{\ldots} \tau_z T_z+J_\perp^{\ldots}(\tau_x T_x+\tau_y T_y)$,
where $\tau_\ga$ are the unity ($\ga=0$) and Pauli ($\ga=x,y,z$) matrices acting in the space of the $E_\ex\pm$ states.
The class (I) model consist of two decoupled two-channel contributions of this kind
and there are no conversion processes $E_{\ex 1}\leftrightarrow E_{\ex 2}$ between them.
In class (II) model the conversion processes $A_1\leftrightarrow A_2$ and $A_{1,2}\leftrightarrow E$ between all IRs are present.
In class (III), the conversion processes $E^\al\leftrightarrow E^\be$ between two 2D IRs of the same type are present as well,
with the structure of the orbital Kondo interactions above.

We note that since the kinetic energy (\ref{eq:H0}) possesses SU(4) rotational symmetry in the channel space $R_\ex$,
the orbital-singlet class (III) model can be transformed to class (I) model without conversion processes $E^\al\leftrightarrow E^\be$
by diagonalizing the matrix $J^{\nu\nu'}$ ($\nu,\nu'=E^\al,E^\be$) of the coupling constants.
Such transformation is, however, not possible in the orbital doublet case,
since this would require diagonalizing 3 matrices $J_0^{\nu\nu'}$, $J_z^{\nu\nu'}$, $J_\perp^{\nu\nu'}$,
at the same time. So the conversion processes $E^\al\leftrightarrow E^\be$
cannot be eliminated in the orbital-doublet class (III) model.

{\em Increase of symmetry.}
We point out the following interesting properties.
In MLG, the local symmetry is increased as $(a,C_{6v})\rtarr (a,D_{6h}=C_{6v}\times C_s)$
and $(b,C_{3v})\rtarr (b,D_{3h}=C_{3v}\times C_s)$ when the IA is moved from the out-of-plane to in-plane position.
This, however, does not lead to qualitative changes in the Kondo models.
In the former case, the channel space $R_\ex=E_1+E_2\rtarr E_1''+E_2''$ still breaks up into two different 2D IRs
and still the class (I) Kondo model is realized, given by the sum of two 2-channel contributions
with generally {\em different} couplings.
This implies, in particular, that the claims~\cite{Sengupta,DellAnna}
of the 4-channel symmetric Kondo effect in the $(a,D_{6h})$ case are unjustified:
group theoretically, this would require the presence of a 4D IR, which is absent in the $D_{6h}$ group).
In the latter case, the channel space $R_\ex=A_1+A_2+E\rtarr A_1''+A_2''+E''$ still breaks up into two 1D and one 2D IRs
and still the class (II) Kondo model is realized.
On the other hand, the increase in symmetry $(c,C_{3v})\rtarr (c,D_{3d})$
upon placing the IA halfway between the BLG sheets does lead to a qualitative change:
the channel space $R_\ex=2E\rtarr E_g+E_u$ breaks up into two {\em different} IRs instead of two IRs of the {\em same type};
as a result, conversion processes vanish and the Kondo model is transformed from class (III) to (I).

To summarize this part, our group-theoretical analysis allow us to make definitive symmetry-based model-independent
conclusions about the structure of the Kondo model, without relying on any specific microscopic details.
In particular, it tells exactly if and how different conduction channels are coupled by the exchange interaction, Tab.~\ref{tab:2}.
We emphasize that generally the channel states $\chi_\eta(\rb)$ can be either pure valley states or mixtures thereof,
as seen from Tab.~\ref{tab:1}. This analysis also resolves the concern~\cite{FV}
that ``valley mixing'' by the impurity potential could
be detrimental for the multichannel Kondo effect:
while valleys can indeed be mixed,
the local symmetry dictates that properly hybridized
valley states act  as independent channels in several instances (Tab.~\ref{tab:2}),
and are also completely equivalent if they belong to one 2D IR.

{\em Feasibility of the channel-symmetric two-channel Kondo effect.}
We now discuss the implications of our results for the realization of the multichannel Kondo effects.
It is well established~\cite{NB,CZ} that the low-energy behavior of the multi-channel Kondo model
is determined by the channel(s) with the largest exchange coupling.
Thus, effectively, the regime of the multi-channel Kondo effect can be realized
only if the common coupling of several channels (belonging to the same multi-dimensional IR)
exceeds all single-channel couplings.
This condition proved to be extremely hard to achieve in practice:
in a ``typical'' band structure with no special properties, some single-channel coupling will usually prevail.

Our findings suggest that, owing to the peculiarities of graphene band structure,
the regime of the dominant two-channel Kondo effect is feasible in MLG and BLG in several cases.

The class (I) Kondo model, which includes the cases $(a,C_{6v}/D_{6h})$ in MLG and $(c,D_{3d})$ in BLG,
consists of two decoupled symmetric two-channel contributions with their own couplings.
Whichever couplings are greater, the low-energy behavior will be dominated by the two channels of that 2D IR.

In the BLG $(d,C_{3v})$ case, if the IA is placed above the center of the carbon hexagon of one layer, as shown in Fig.~\ref{fig},
the atomic orbitals of the 2D IR $E$ channels, located on the $\At$ sublattice in that nearby layer, are much close to the impurity
than those of the 1D IR $A_{1,2}$ channels, located on $B$ sublattice in the remote layer.
Thus the hybridization with the $E$ channels is likely to be considerably greater than with $A_{1,2}$ channels.
This should result in a greater exchange couplings $J^{EE}_{\ldots}>J^{A_{1,2}A_{1,2}}_{\ldots}$ and
the low-energy behavior will be dominated by the two $E$ channels.

In these cases, in the low-energy regime, the symmetric two-channel Kondo models
(\ref{eq:HJ1low}) and (\ref{eq:HJ2low})
will be realized in the orbital singlet and doublet cases, respectively, in the subspace of the two dominant channels $E_\ex\pm$.

{\em Conclusion and Outlook.}
In summary, motivated by the prospect of realizing multichannel Kondo effects,
we performed a general group-theoretical classification of the Kondo models in MLG and BLG in the vicinity of the CNP
for all placements of the IA with either 3- or 6-fold rotational symmetry.
We found six possible classes of the four-channel Kondo models, summarized in Tab.~\ref{tab:2}.
We argued several possibilities for realizing the channel-symmetric two-channel Kondo effect,
described by the models (\ref{eq:HJ1low}) and (\ref{eq:HJ2low}),
which are known to exhibit non-Fermi-liquid behavior in a number of regimes but have proven challenging to realize in practice.
Our findings thus open prospects for the observation of the multichannel Kondo physics
in MLG and BLG, which could be pursued experimentally using the local probes, such as scanning tunneling microscopy, or transport measurements.

{\em Acknowledgement.}
This work was supported by the U.S. DOE under contract No. DE-FG02-99ER45790
and by the NSF grant No. DMR-0906943.

\begin{widetext}

\section{Supplementary Material: Radial channels in graphene}

In MLG, the solution of the eigenvalue problem $\hh_\la^\text{MLG} \psi = \e \psi$, $\la=K,K'$, in the polar coordinates $\rb=r(\cos\vphi,\sin\vphi)$
is provided by the eigenfunctions
\beq
    \psi_{\e K j_z}(\rb) = \Nc_\e\lt( \begin{array}{c}
        J_{|j_z-\fr12|}(kr) \ex^{\ix\lt(j_z-\fr12\rt)\vphi}
        \\ \ix\,\sgn\e\,\sgn j_z\, J_{|j_z+\fr12|}(kr) \ex^{\ix\lt(j_z+\fr12\rt) \vphi}  \end{array}\rt)_{AB},
    \mbox{ }
            \psi_{\e K' j_z}(\rb) = \psi_{\e K j_z}^*(\rb) \mbox{ (MLG)}
\lbl{eq:psiMLG}
\eeq
Here, $J_{...}(kr) $ are the Bessel functions and $k=|\e|/v$.
The eigenstates are characterized by the energy $\e$, valley $\la=K,K'$, and angular momentum $j_z$ quantum numbers;
the latter takes half-integer values, $j_z=\pm\fr12,\pm\fr32, \ldots$.

In BLG, the solution of the eigenvalue problem $\hh_\la^\text{BLG} \psi = \e \psi$
in the polar coordinates is provided by the eigenfunctions
\beq
    \psi_{\e K j_z}(\rb) = \Nc_\e\lt( \begin{array}{c}
        J_{|j_z+1|}(kr) \ex^{\ix(j_z+1)\vphi}
        \\ -\sgn\e\, J_{|j_z-1|}(kr) \ex^{\ix(j_z-1) \vphi}  \end{array}\rt)_{\At B},
    \mbox{ }
                \psi_{\e K' j_z}(\rb) = \psi_{\e K j_z}^*(\rb) \mbox{ (BLG)}
\lbl{eq:psiBLG}
\eeq
where $|\e|=\fr{k^2}{2m}$ and the angular momentum $j_z=0,\pm 1,\pm 2,\ldots$ takes integer values.

The normalization of the states as
\[
    \int\dx^2\rb\,\psi^\dg_{\e \la j_z}(\rb) \psi_{\e'\la' j_z'}(\rb) = 2\pi\de(\e-\e') \de_{\la\la'} \de_{j_z j_z'}
\]
yields for the constant
\[
    \Nc_\e=    \sq{\pi\nu_\e},
\mbox{ where }
    \nu_\e=\lt\{\begin{array}{ll}
        \fr{|\e|}{2\pi v^2} &\mbox{(MLG)} \\
        \fr{m}{2\pi} &\mbox{(BLG)} \end{array}\rt.
\mbox{ is the density of states. }
\]

The expansion of the electron field operator
in terms of the annihilation operators $\ch_{\e\la j_z\sig}$ of the states (\ref{eq:psiMLG}) or (\ref{eq:psiBLG}) reads
\beq
    \psih_{\la\sig}(\rb) = \sum_{j_z} \int\fr{\dx\e}{2\pi}\, \psi_{\e\la j_z}(\rb) \ch_{\e\la j_z\sig} \mbox{ } (\la=K,K', \mbox{ } \sig=\ua,\da),
\lbl{eq:psiexp}
\eeq
The operators $\psih_{\la\sig}(\rb)$ are spinors in the sublattice space, $AB$ for MLG and $\At B$ for BLG,
which are convenient to join into the eight component spinor
\[
    \psih(\rb)= \lt(\begin{array}{c} \psih_{K\ua}(\rb) \\ \psih_{K\da}(\rb) \\ \psih_{K'\ua}(\rb) \\ \psih_{K'\da}(\rb) \end{array}\rt)
\]
in the valley-sublattice-spin space.

The kinetic energy has the form
\beq
    \Hh_0 = \sum_{\la j_z\sig}\int_{-\infty}^\infty \fr{\dx\e}{2\pi} (\e-\e_F) \ch_{\e\la j_z\sig}^\dg \ch_{\e\la j_z\sig}.
\label{eq:H0SM}
\eeq

The most general form of the exchange Hamiltonian, invariant under spin rotations, reads
\beq
    \Hh_J  = \psih^\dg(0) (\hat{\Jc}^\rho+\hat{\Jc}^\sig \sigb \Sb) \psih(0),
\lbl{eq:HJSM}
\eeq
where $\hat{\Jc}^{\rho,\sig}$ are $4\times 4$ matrices in the valley-sublattice space
(containing the impurity orbital isospin operators in the doublet case).
Only the states (\ref{eq:psiMLG}), (\ref{eq:psiBLG}) with nonvanishing weight at $r=0$ couple to the point impurity.
This is the case only for $j_z=\pm\fr12$ states in MLG,
\[
    \psi_{\e \la j_z=\fr12}(0)=\Nc_\e\lt( \begin{array}{c} 1 \\ 0 \end{array}\rt)_{AB},
\mbox{ }
    \psi_{\e \la j_z=-\fr12}(0)=\Nc_\e\lt( \begin{array}{c} 0\\ \ix\,\sgn\e \sgn \la \end{array}\rt)_{AB}
    \mbox{ (MLG),}
\]
and only for $j_z=\pm 1$ states in BLG,
\[
    \psi_{\e\la j_z=-1}(0)=\Nc_\e\lt( \begin{array}{c} 1 \\ 0 \end{array}\rt)_{\At B},
\mbox{ }
    \psi_{\e\la j_z=1}(0)=\Nc_\e\lt( \begin{array}{c} 0\\ -\sgn\e \end{array}\rt)_{\At B} \mbox{ (BLG).}
\]
Importantly, for given $j_z$, the wavefunctions of these states reside on either one of the sublattices at $r=0$.
Thus, in both MLG and BLG, there is exactly one radial channel per valley-sublattice,
and one can identify the angular momenta and sublattice degrees of freedom:
\beq
    j_z=\fr12 \leftrightarrow A \mbox{ and } j_z=-\fr12 \leftrightarrow B     \mbox{ (MLG),}
\lbl{eq:jzMLG}
\eeq
\beq
    j_z=-1 \leftrightarrow \At \mbox{ and } j_z=1 \leftrightarrow B    \mbox{ (BLG).}
\lbl{eq:jzBLG}
\eeq

Substituting the expansion (\ref{eq:psiexp}) into Eq.~(\ref{eq:HJSM}), we obtain
\beq
    \Hh_J  = \int \fr{\dx\e}{2\pi} \fr{\dx\e'}{2\pi} \ch_\e^\dg [\Jh^\rho(\e,\e')+ \Jh^\sig(\e,\e') \sigb\Sb] \ch_{\e'},
\lbl{eq:HJSM2}
\eeq
where
\[
    \Jh^{\rho,\sig}(\e,\e') = \pi\sq{\nu_\e\nu_{\e'}}\hat{\Jc}^{\rho,\sig}
\]
and we join the operators of the eight states (for a given $\e$)
that couple to the impurity into the spinor
\[
    \ch_\e=\lt(\begin{array}{c}\ch_{\e\ua} \\ \ch_{\e\da}\end{array}\rt),
\mbox{ }
    \ch_{\e\sig} =\lt(\begin{array}{c} \ch_{\e KA\sig}\\ \ch_{\e KB\sig}\\ \ch_{\e K'A\sig}\\ \ch_{\e K'B\sig}\end{array}\rt) \mbox{ in MLG and }
    \ch_{\e\sig} =\lt(\begin{array}{c} \ch_{\e K\At\sig}\\ \ch_{\e KB\sig}\\ \ch_{\e K'\At\sig}\\ \ch_{\e K'B\sig}\end{array}\rt) \mbox{ in BLG.}
\]
and use the identifications (\ref{eq:jzMLG}) and (\ref{eq:jzBLG}) of $j_z$ and sublattice degrees of freedom.

Equations~(\ref{eq:H0SM}) and (\ref{eq:HJSM}) present the Kondo Hamiltonian $\Hh_0+\Hh_J$ in the basis of the operators $\ch_{\e \la j_z\sig}$.
This Hamiltonian can be rewritten in terms of the chiral 1D electrons introduced as
\[
    \psih(s)=\int \fr{\dx\e}{2\pi} \ex^{\ix \e s} \ch_\e,
\]
where $s<0$ and $s>0$ regions correspond to the incoming and outgoing radial waves
(we preserve the same notation for fields $\psih(\rb)$ in 2D and $\psih(s)$ in 1D, since this should not lead to confusion).
Leaving only the four channels that couple to the impurity, the kinetic energy (\ref{eq:H0SM})
takes the form of Eq.~(\ref{eq:H0}) of the Main Text in this basis.
The exchange interaction (\ref{eq:HJSM}) takes the local form of Eq.~(\ref{eq:HJ}) of the Main Text in terms of $\psih(s)$
if $\Jh^{\rho,\sig}(\e,\e')$ can be treated as energy-independent.
This holds exactly in BLG (since its density of states $\nu=\fr{m}{2\pi}$ is $\e$-independent)
and holds in MLG in the weak-coupling limit, where the value $\Jh^{\rho,\sig}=\Jh^{\rho,\sig}(\e_F,\e_F)$ at the Fermi level is
taken.

\section{Supplementary Material: Details of the derivation of the Kondo models}

Here, we provide some details of the group-theoretical derivation of the Kondo models summarized in Tab.~\ref{tab:2} of the Main Text.

{\em Impurity orbital singlet class.}
In the orbital singlet class, $R_\ix=A_\ix$,
since the square of any 1D IR is a unity IR, $R_\ix\times R_\ix^\dg=A_1/A_1'/A_{1g}$,
the transformation properties of
$
    \Jh^{\rho,\sig} \sim R_\ex \times R_\ex^\dg
$
are determined by those of the conduction electrons.

If in the decomposition of $R_\ex$ (Tab.~\ref{tab:1}) all IRs are of different type,
then only the squares of each IR contain a unity IR and produce an invariant,
which is a unity matrix in the subspace of the IR.
As a result, there are no processes transferring electrons between different channels, neither within one 2D IR nor between
different IRs: $\Jh^{\rho,\sig}$ are diagonal matrices
and within each 2D IR both channels have the same coupling constant, protected by symmetry.
This is so for classes (I) and (II) but not for (III), i.e., in all cases we consider, except for the BLG $(c,C_{3v})$ case.
Thus, the orbital-singlet class (I) and (II) Kondo models contain two and one symmetric two-channel contributions, respectively,
decoupled from other channels, a prerequisite for the realization of the two-channel Kondo effect.

On the other hand, in class (III), the conduction sea $R_\ex=2E=E^\al+E^\be$ consists of two 2D IRs $E^{\al,\be}$ of the same type $E$
(the labels $\al,\be$ are used to distinguish between the two subspaces).
As a result, the decomposition $R_\ex\times R_\ex^\dg= 4(A_1+A_2+E)$ contains four invariants ($4A_1$):
two correspond to the symmetric coupling within each 2D IRs and
and two,described by one {\em complex} coupling constant $J^{E^\al E^\be}$,
correspond to ``conversion'' processes $E^\al\leftrightarrow E^\be$, whereby the conduction electrons are transferred between two IRs.

{\em Impurity orbital doublet class.}
In the orbital doublet class, $R_\ix=E_\ix$, for any case of the impurity placement,
the product $R_\ix\times R_\ix^\dg = A_1[T_0]+A_2[T_z]+E[T_{x,y}]$
decomposes into two 1D and one 2D IRs,
whose bases are formed by the impurity isospin operators 
indicated in brackets. Likewise, for any given 2D IR $E_\ex$ in $R_\ex$,
the bases of the IRs in the decomposition
$E_\ex\times E_\ex=A_{1}[\tau_0]+A_{2}[\tau_z]+E[\tau_{x,y}]$
are formed by the unity and Pauli matrices $\tau_\ga$, $\ga=0,x,y,z$ acting in the space of $E_\ex\pm$
conduction electron states. (Here, the MLG $(b,C_{3v})$ case is provided as an illustrative example,
all other cases are shown in the fourth column of Tab.~\ref{tab:1}).

Evaluating the contributions to the product (\ref{eq:P}),
we find that, for any class (I), (II), or (III) and any 2D IRs $E_\ex$ and $E_\ix$,
the product $
    (E_\ex\times E_\ex)\times (E_\ix\times E_\ix)
$
contains three invariants,
which give rise to an anisotropic orbital Kondo interaction
$J_0^{\ldots} \tau_0 T_0+J_z^{\ldots} \tau_z T_z+J_\perp^{\ldots}(\tau_x T_x+\tau_y T_y)$
between the impurity isospin and the conduction electrons in two $E_\ex$ channels,
in both charge ($\Jh^{\rho}$) and spin ($\Jh^{\sig}$) sectors.

Further, the product
$(E_{\ex1}\times E_{\ex2})\times(E_\ix\times E_\ix)$ with different 2D IRs $E_{\ex1}$ and $E_{\ex2}$
contains no invariants.
Thus, in class (I),
there are no conversion processes $E_{\ex 1}\leftrightarrow E_{\ex 2}$
and the Kondo model consists of two decoupled two-channel contributions.
In class (II),
the products $(A_1\times A_2)_\ex\times(E\times E)_\ix$ and $(A_{1,2}\times E)_\ex\times(E\times E)_\ix$
contain invariants, and hence conversion processes $A_1\leftrightarrow A_2$, $A_{1,2}\leftrightarrow E$ between all IRs are present.
In class (III), the invariants of $(E\times E)_\ex\times (E\times E)_\ix$
also result in conversion processes $E^\al\leftrightarrow E^\be$,
with the structure of the orbital Kondo interactions above.

\end{widetext}


\begin{thebibliography}{99}
\bibitem{TR}
$\Hh_J$ must also remain invariant under the time reversal;
we find, however, that the time reversal symmetry does not impose any additional constraints.
\end{thebibliography}
\end{document}